\documentclass[sigconf]{acmart}

\usepackage{booktabs} 

\begin{document}
	\title{OmniArt: Multi-task Deep Learning for Artistic Data Analysis}

\author{Gjorgji Strezoski}
\affiliation{%
  \institution{University of Amsterdam}
  \city{Amsterdam} 
  \state{The Netherlands} 
}
\email{g.strezoski@uva.nl}

\author{Marcel Worring}
\affiliation{%
	\institution{University of Amsterdam}
  \city{Amsterdam} 
  \state{The Netherlands} 
}
\email{m.worring@uva.nl}

\renewcommand{\shortauthors}{G. Strezoski et al.}

\begin{abstract}
	Vast amounts of artistic data is scattered on-line from both museums and art applications. Collecting, processing and studying it with respect to all accompanying attributes is an expensive process. With a motivation to speed up and improve the quality of categorical analysis in the artistic domain, in this paper we propose an efficient and accurate method for multi-task learning with a shared representation applied in the artistic domain. We continue to show how different multi-task configurations of our method behave on artistic data and outperform handcrafted feature approaches as well as convolutional neural networks. In addition to the method and analysis, we propose a challenge like nature to the new aggregated data set with almost half a million samples and structured meta-data to encourage further research and societal engagement. 
\end{abstract}

\maketitle

\section{Introduction}

Art applications like WikiArt \footnote{https://www.wikiart.org/}, Europeana \footnote{https://www.europeana.eu/portal/en}, ArtUk \footnote{https://www.artuk.org/}, WGA \footnote{https://www.wga.hu} and Google Art Project \footnote{https://www.google.com/culturalinstitute/beta/} have aggregated a diverse set of art collections together with basic meta-data and made them public using the web. Museums on the other hand expose a much larger part of the meta-data and structure based on a museum-centric point of view. Usually these collections contain far richer, expertly designed meta-data than the ones found in on-line art applications. Using the combination of accurate meta-data with quality photographic reproductions of the items in their collections, museum-centric data are ideal for analysis as they allow for deeper art exploration from multiple perspectives. 

In such museum-centric collections there is a gap between the available information and the data contained in the annotations. Often times, the semantic meta-data available for a specific piece of art is relayed differently in different use case scenarios. This poses a problem in searching and indexing these collections. \textit{Shreiber et al.} made a significant step forward into bridging the gap in attributing meta-data to artworks in a standardized fashion \cite{schreiber2008semantic}. They collected over 200,000 art samples from various collections on top of which they created a vocabulary for describing artworks. Using this vocabulary, they mapped the existing meta-data with RDF relations to other entities in the Linked Open Data cloud. Doing this effectively provided standardized annotations and vastly expanded the meta-data available with semantic context.

\begin{figure}

  \includegraphics[width=\linewidth]{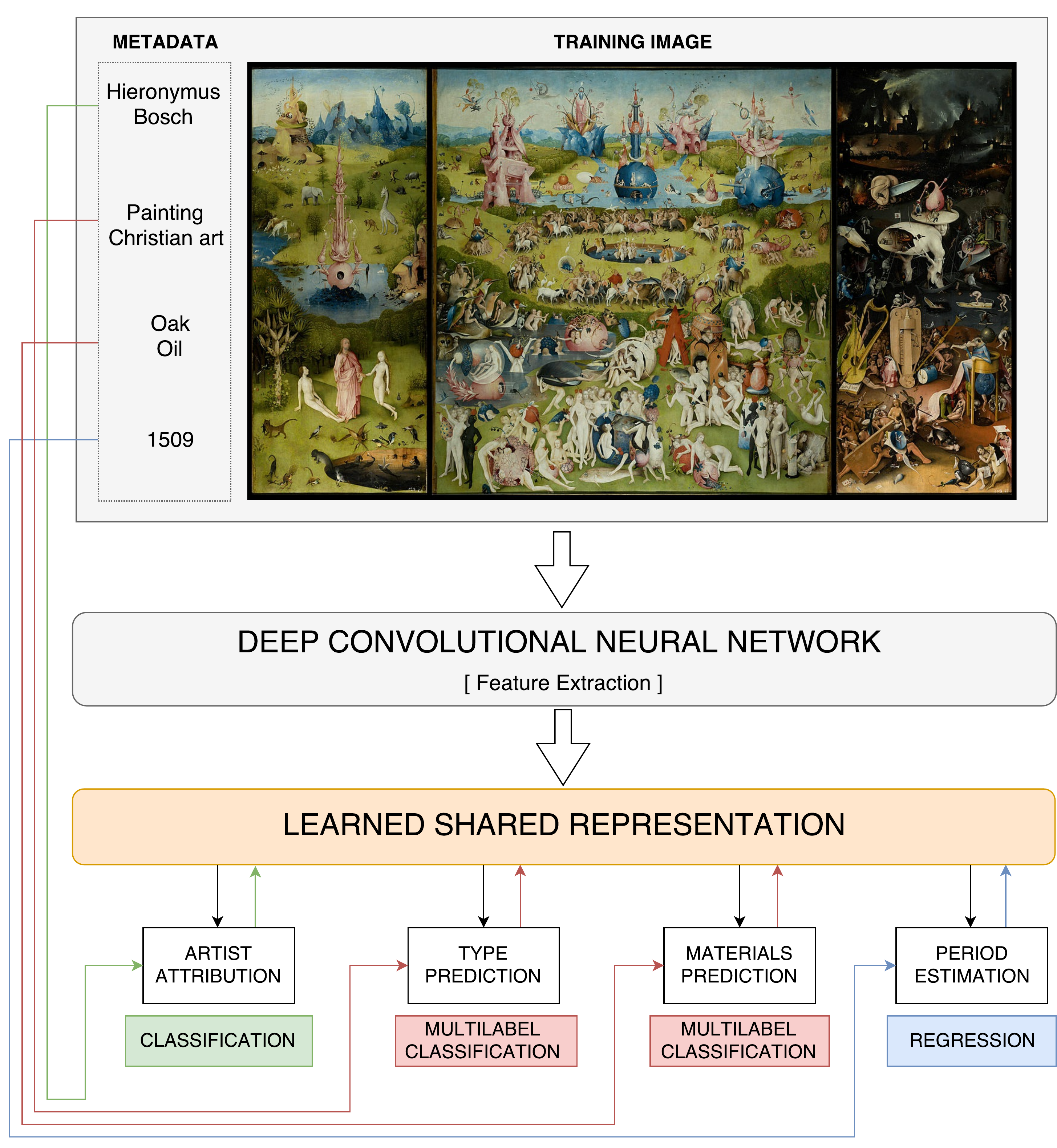}
  \caption{OmniArt - High level method overview\label{fig:figure_1}}
\end{figure}

However, even without the linked data expansion, cultural heritage is in general an outstandingly knowledge rich domain. In artistic paintings for example, most of the artworks have a known artist, style, year of creation, materials, geographical origins and even very detailed textual descriptions about their content. With specific methods from material science, chemical compositions of color can be extracted together with canvas thread patterns in paintings \cite{yang2015quantitative}. Information is available on an even finer scale for these data sets. Using high resolution photography and x-rays \cite{klockenkamper2000analysis, chung1999industrial} we are able to see the illusive details and generate more insight than ever before \cite{pouyet20152d}. Each mentioned chunk of information presents a different challenge for scientists. With the growing amount of this museum-centric, context rich data \footnote{http://www.metmuseum.org/art/collection} \footnote{https://www.rijksmuseum.nl/en/api}, the need for efficient tools for exploring and analyzing art collections is ever more imminent. For this work we focus on a sub-domain of this data pool, which is the raw photographic reproductions of artworks in combination with the textual meta-data provided by the museums.

Our focus area in meta-data contains attributes for which we hypothesize are semantically linked. Having multiple types of attributes, creates the possibility of having separate tasks for each type giving this research a multi-task nature. The most time efficient way to tackle multiple tasks is to do it simultaneously, which is why in this paper we propose a multi-task deep learning method with shared data representation between tasks depicted in Figure \ref{fig:figure_1}. Creating a shared representation allows us to exploit the semantic entanglement between the different tasks. With encouraging scientific progress in bridging the aforementioned information gap in mind, we propose a challenge to the tasks addressed in this paper.

Challenges have been repeatedly proven as a good catalyst in stimulating a community to contribute to a cause. A shining example of a successful challenge in multimedia is The Multimedia Grand Challenge \cite{snoek2011academia} that has been bringing commercial industrial needs to the attention of scientific researchers since 2009. In computer vision, the ImageNet Large Scale Visual Recognition Competition (ILSVRC) gave way for some of the largest breakthroughs. One such challenge for art, combining the information abundant artistic data, while providing a museum-centric perspective, was introduced in 2014 as The Rijksmuseum challenge by \textit{Mensink et al.} \cite{mensink2014rijksmuseum}. The Rijksmuseum challenge is composed of four separate tasks, namely: \textbf{artist} attribution, predicting the \textbf{art-type}, predicting the \textbf{materials} used and predicting the \textbf{creation year} of specific artworks in the collection. In 2014 this data set contained close to 120,000 entries of various photographic reproductions of the artworks in the Rijksmuseum's on-line collection. A single artwork in the challenge is described by a number of attributes like artist, period of creation, materials used, type of artwork and so on. The works of art in this data set date from ancient times to the late 19th century and are attributed to over 6000 individual artists.

Even though artistic data provides multiple ways of analyzing it, most current methods for artistic data analysis \cite{bar2014classification, karayev2013recognizing, elgammal2015quantifying, Roman-Rangel2016, gatys2016image, guccluturk2016convolutional} address each task individually. It would be more efficient to tackle this multi-task problem in a single end-to-end system, as this way the training time required for multiple models is reduced significantly. In some cases the multi-task learning environment can even improve classification performance if there is a correlation between the categories in each of the tasks. Work done by \textit{Kokkinos et al.} \cite{DBLP:journals/corr/Kokkinos16} successfully show the benefits of a deep model adapted for solving multiple tasks at once. Their architecture attempts to perform multiple computer vision tasks with one propagation of the input data through the model, which partly inspired our work. 

Using our proposed method, called OmniArt we report state-of-the-art results on The Rijksmuseum Challenge from 2014 and propose a new challenge with an expanded a better structured data set. Upon acceptance, we will make the challenge publicly available with the data set, trained models and evaluation engine to stimulate further development.

In this work we elaborate on the following contributions:

\begin{itemize}
  \item  We propose an efficient and accurate multi-task end-to-end method for learning a shared representation of the input data with respect to all tasks.
  \item  We offer a museum-centric data set with more than 430,000 samples with updated meta-data dubbed The OmniArt Challenge to stimulate engagement, encourage new research and maximize societal impact.   
  \item  We report state-of-the-art results on The Rijksmuseum Challenge from 2014 and The OmniArt Challenge with significantly shorter training duration.
\end{itemize}

The rest of the paper is structured as follows: Section 2 contains the related work from a both a multi-task learning perspective and general artistic data analysis. Section 3 introduces the proposed method and the logic behind it. Section 4 is about the experimental setup, the datasets used for training and testing and experimental results. In the final section we give our concluding remarks with qualitative findings from the analysis we performed.

\section{Related Work}
Related work in this area can be divided into two segments namely artistic data analysis and multi-task learning.

\subsection{Artistic Data Analysis}

As early as 1979, \textit{J. Rush} \cite{rush1979acquiring} concluded that experiences with individual instances of art from a particular artist can lead to the ability of identifying works from the same artist which have not been seen before. While a pure visual experience with samples from an artist efficiently taught the subjects to recognize such never before seen artworks, performance experienced a significant boost when other contextual information was presented in combination with the original image. With added context, possible sources of confusion were eliminated and recognition performance spiked. \textit{Jonson et al.} \cite{johnson2008image} performed a detailed analysis of brush-strokes in the work of Van Gogh using Gabor, Complex and D4 wavelets in combination with Support Vector Machines (SVM) and thresholding. This analysis has been done on a very small scale of just 101 images with full resolution reproductions as input. They conclude that brush-stroke analysis is helpful in artist attribution but it also depends on a lot of external factors like the canvas degradation and pigment loss. For a large scale accurate analysis, artworks need a scaled down representation with minimal information loss. Different from \cite{rush1979acquiring, johnson2008image} we propose doing such categorization on a large scale.

As one of the larger artistic datasets, the Rijks'14 data set was introduced by \textit{Mensink et al.} \cite{mensink2014rijksmuseum} in 2014 for The Rijksmuseum Challenge. At this point baseline scores for the data set were computed with the help of opponent and intensity SIFT features \cite{van2010evaluating} represented in Fisher vectors. For classification they utilized the liblinear SVM library \cite{fan2008liblinear}. On the same data set, \textit{Van Noord et al.} \cite{van2015toward} performed artist attribution using their own subsets with a convolutional neural network named PigeoNET. Their implementation features five convolutional and three stacked fully connected layers like Alexnet. Performance with artist attributions is reported on subsets with three sources of variation: 1) heterogeneity versus homogeneity, 2) number of artists in the set and 3) number of artworks per artist. Conclusions drawn from this research state that the performance of the model is proportional with the number of samples per class - more samples per class equals better attribution capabilities. Additionally, when the model is trained on a single type of artwork (for example only prints), performance increases since the model does not have to deal with big variations between artworks from the same artist. Although, similar artwork types improve the ability to learn better features, sometimes it can make classification confusing due to the sample similarity. \textit{Van Noord et al.} \cite{van2015toward} present an extensive analysis into artist attribution, but make no use of the other meta-data (period, materials, types...) which we exploit and prove beneficial for determining the attributes of an artwork. 

Another large body of artistic data is the WikiArt data set. Multiple artistic data analysis approaches \cite{bar2014classification, elgammal2015quantifying, karayev2013recognizing, saleh2015large} have been tested on WikiArt as it has quality annotations for artists, periods and art types. However, due to missing the material information about the artworks we only contain a subset of the WikiArt data in the OmniArt challenge. The Picasso data set used in \cite{ginosar2014detecting, westlake2016detecting} for people detection features 218 Picasso paintings, and most of them are already included as a subset of the current version of the new data set.

When it comes to art, tangible information like artists and periods is only one piece of the puzzle. Style also plays a significant role in identifying the origins of an artwork. In 2016 \textit{Gatys et al.} \cite{gatys2016image} proposed a style transfer method using an energy-minimization point of view. They use a pre-trained convolutional neural network \cite{DBLP:journals/corr/SimonyanZ14a} as a feature extractor for both the style origin image and the image the style should be transfered to. Capturing such detail and transferring it in a meaningful fashion shows that quality information can be extracted from artistic data using convolutional neural networks. Another very recent generative approach to artistic data is presented in \cite{CycleGAN2017} where \textit{Zhu et al.} effectively transfer style, enhancements and transfigurations on images without pairing using Generative Adversarial Training. 

\subsection{Multi-task Learning}

Multi-task Learning is a paradigm of inductive transfer where the goal is to improve generalization performance by leveraging the domain-specific information of the training data in the related tasks \cite{caruana1998multitask}. With the datasets becoming larger and larger, the idea of maximally exploiting each pass-through of data becomes rather attractive and approaches analyzing multiple aspects of the data in one go are becoming increasingly popular. Given the merits of multi-task learning, this paper addresses art data analysis from a multi-task point of view in a categorization setting.

\textit{Kokkinos} in \cite{DBLP:journals/corr/Kokkinos16} introduces a convolutional neural network architecture that jointly handles visual features on different levels named UberNet. In his work he generates object boundaries, saliency maps, semantic segmentation, surface normals, and detection, in a single forward pass of an image through the model. Ubernet is part of the motivation behind our work as it uses a similar end-to-end paradigm for a multi-task problem. While it uses a clear separation between tasks, Ubernet does not allow for significant information sharing between tasks other than a joint loss affecting all layers below a specific output. Nevertheless, sharing information between different task representations proves to be beneficial for the model in training as \textit{Misra et al.} conclude in \cite{misra2016cross}. A \textit{stitching layer} as they address it, provides a shared structure of units that combines the activations from multiple networks into one end-to-end trainable model. In natural language processing, the multi-task approach to deep learning has proven beneficial as well. \textit{Liu et al.} \cite{liu2015representation} perform multiple-domain classification on texts using multiple shared layer representations. Spanning different domains, multi-task learning can be done either with explicit \cite{liu2015representation, misra2016cross} or implicit information sharing \cite{DBLP:journals/corr/Kokkinos16}.

Recent studies have shown that information sharing between tasks can be beneficial \cite{yang2014unified} for action detection \cite{DBLP:journals/corr/ZhuN16c}, zero-shot action recognition \cite{DBLP:journals/corr/XuHG16}, human pose estimation \cite{wang2016human} and adaptive visual feedback generation for facial expression improvement \cite{Kaneko:2016:AVF:2964284.2967236}. Current methods use different layer depths to address tasks with varying complexity or use multiple inputs in their models so the different tasks have suitable features for training the classifier/regressor in the final block. While in our approach, we use the same features for each task and apply task specific scaling and weighting depending on the task's nature. An added benefit of our approach is that if there is even a slight correlation between the targets of the different tasks, it can improve the overall models performance. In our method we aim to learn a semantic link between tasks and use that insight to simultaneously predict multiple attributes about an artwork in an efficient and accurate fashion.

\section{Method}

\begin{figure*}
  \includegraphics[width=\linewidth]{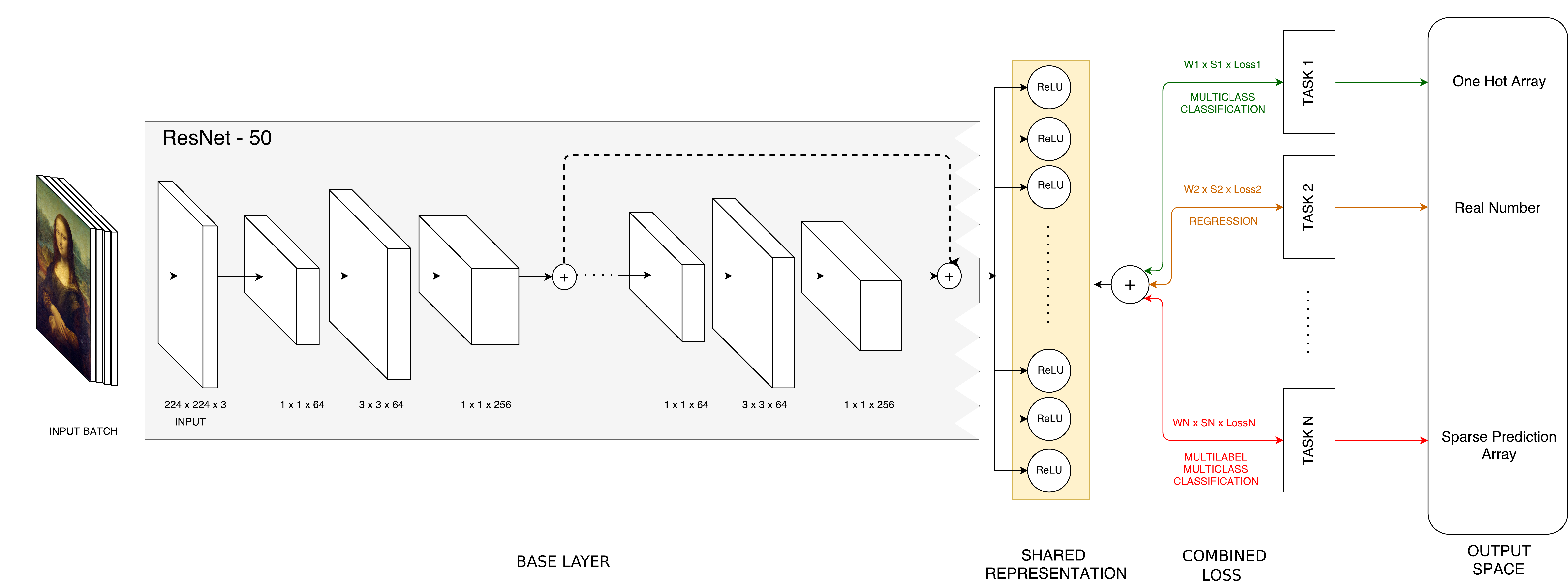}
  \caption{Proposed multi-task deep convolution neural network architecture with a shared representation layer - OmniArt\label{fig:omniart_artchitecture}}
\end{figure*}

Training separate models for each of the tasks in this data set is a computationally inefficient and time-consuming process. As is the case with multi-task datasets, the images propagating through the model, in the forward pass, are identical for each task. A difference can only be observed in the back-propagation from the final classification/regression block due to the different label spaces, dimensions and loss types. Moreover, it is common for these types of tasks that there is correlation between the different label types, influencing the outcome of a certain prediction. This means that a certain artwork with a known period of creation and artwork type significantly narrows down the list of possible artists for the artist attribution task. A mathematical interpretation of this example is shown with a simple conditional probability in equation 1 where
\begin{math}
T_{1}
\end{math}
represents the artwork belonging to a particular artist, while 
\begin{math}
T_{2}
\end{math}
and
\begin{math}
T_{3}
\end{math}
correspond to the the period of creation and the type of material used.
\begin{equation}
P(T_{1}|T_{2},T_{3})=\frac{P(T_{1}\bigcap T_{2}|T_{3})}{P(T_{2}|T_{3})}
\end{equation}

A real world example of this type of correlation would be a painting which has a creation period of \textit{1635} and a type of \textit{oil on canvas}. The chances of this painting being a \textit{Van Gogh} are close to none, because \textit{Van Gogh} was not born until 1853. It would more likely be a \textit{Rembrandt} since he was active in that time period. Therefore we hypothesize that there is a semantic entanglement between the different attributes in artistic data. 

\subsection{Method overview}

In this paper we propose a multi-task learning method for learning a shared representation based on the semantic entanglement between multiple attributes of artistic data. As depicted in Figure \ref{fig:omniart_artchitecture}, our method consists of a base layer block for feature extraction, a shared representation block, a combined loss layer where the loss from all tasks is aggregated and separate evaluation blocks per task. We perform loss aggregation in a sum with weighing and scaling the separate losses depending on the type of task they originate from and the impact of that particular attribute to the overall performance. 
Using this method we improve discriminative performance on every task, reduce training and testing duration because of the one-time data set traversal.

\subsection{Combined loss layer}

We propose a multi-task convolutional neural network which learns a shared representation of the artworks with respect to multiple types of accompanying meta-data. For each of the meta-data attributes we create separate tasks and assign a separate classification/regression block in the model, each with their own loss function. A natural way to efficiently propagate these gradients through our model and have them all influence the training properly is in a sum. In equation 2,  
\begin{math}
L_{t}
\end{math}
is the combined loss for all of the tasks, while 
\begin{math}
L_{i}
\end{math}
presents the loss per individual task. Parameters 
\begin{math}
w_{i}
\end{math}
and
\begin{math}
s_{i}
\end{math}
are representing the task specific loss weight matrix assigning a different weight to each task and the task-specific scale factor respectively.

\begin{equation}
L_{t}=\sum_{i=1}^{n}w_{i}s_{i}L_{i}
\end{equation}

It is important to note that this way of combining loss functions only works if they share a layer with trainable parameters. Sharing such a layer would imply for them to get input from the same level. In terms of loss functions, for classification tasks like artist attribution we use a categorical cross-entropy loss with a softmax function. For regression we apply scaled mean absolute loss and an interval accuracy with a tolerance of $\pm 50$ years. Since interval accuracy intrinsically implies classification, we utilize the mean absolute loss for training the regression block. Two of the tasks in this challenge are multi-label classification tasks, so we utilize a binary cross-entropy loss function over the sparse labels with sigmoid activation. 

While this type of loss aggregation comes to mind naturally in a scenario like this, the effect and scale of the loss value has to be taken into account \cite{sculley2010combined}. In fact, the loss values generated by the regression tasks prove to be at least 10 times larger then the ones generated in the classification tasks or multi-label tasks. This influences the learning negatively as the adjustment to the shared representation is most influenced by the regression tasks, while the importance of the classification losses is diminished. In order to even the influence out and make the assigned task weights reliable again, we scaled down the regression (mean absolute) loss by a factor of ten manually 
\begin{math}
(s_{period} = 10) 
\end{math}
when back-propagating. The specific scale factor can be determined through monitoring the loss values in the validation phase. 

\subsection{Shared representation layer}

Since we are using a deep model as a feature extractor, we limit the back-propagation effects to only the additional layers (outputs per task and shared layer). In turn, we only train a small amount of parameters compared to the total number of parameters contained in the model because:

\begin{itemize}
  \item We do not have enough labeled data to effectively adjust the filters in the deep model from scratch (for example Resnet-50 has 25,583,592 trainable parameters without the top output layers).
  \item Training only the final output blocks speeds up the whole process while still learning a good representation.
  \item The data dimensions are more manageable and the training effects are easier to study.
\end{itemize}

Given that between each of the classes in the different tasks
\begin{math}
(T_{i})
\end{math}
there exists a joint probability, the shared layer is a joint representation of the data with respect to each task. We experimentally determined the shared layers configuration in terms of activations and amounts of hidden units. Different activation functions like \textit{Tanh} \cite{lecun2012efficient} and \textit{Sigmoid} \cite{lecun2012efficient} do not promote sparsity in our representation like \textit{ReLU} \cite{nair2010rectified} does. We believe that the sparsity factor plays a key role in the information absorption of the shared layer. The number of hidden units in the shared layer is dependent on the number of output targets per task and the diversity in the data itself. From a learning point of view this is expected, as more targets would require more trainable parameters for learning a valid representation and vice versa.

Another point of view, derived from the experiments in stage one of this work would suggest that not all tasks require the same depth of output. Experimental results show that the type and period prediction tasks can be efficiently addressed with a shallower architecture. This would also imply a speed up in training times and also reduce memory consumption as the number of trainable parameters would decrease both in a fine-tune and from-scratch setting. However, dispersing the output layers in different depths of the model implies that we cannot model a joint loss influencing the combined data representation for all the tasks at once in a shared representation layer. 

Since this layer is not an output of the network it can also be used as a high-level feature extraction point. If the model performs well on each task, the features extracted at this point would be a valid representative of the input data. With lower dimensions because of the limited number of units in the shared layer, these features would be preferred when memory and computing capacity is limited. 

\subsection{Base Layer}

Our method is essentially agnostic to the choice of a feature extractor as the base for the shared representation because the features that are used for prediction and evaluation are the ones learned in the shared layer. With the success of deep models in visual recognition tasks, we experimented with a number of different deep architectures like VGG-16, VGG-19, Inception V2 and ResNet-50 as feature extractors. Our experimental results suggest that the ResNet-50 model generated the best base features, so we assigned as the feature extractor in our method.

Features are extracted from the last layer before the classifier and propagated through a shared representation layer to a different evaluation blocks for each task. Back-propagation of the combined loss modifies the features in the shared representation layer with respect to every task. 

\section{Experiments}

Through the course of our experiments we are aiming to answer the following questions:

\begin{itemize}
  \item Which deep model performs best as a base feature extractor in artistic datasets?
  \item Does learning a representation on multiple interconnected tasks improve overall predictive performance and if so, how?
  \item How do different types of tasks (classification, regression, multi-label) affect each other in a combined setting?
  \item Which parameters work best in a shared representation setting when the tasks are of different types?
  \item Can the shared representation learn the semantic connections between the tasks and generate qualitative insight?
\end{itemize}

Our experimental setup features two stages. In the first stage we evaluate the performance of individual models for each task in the Rijksmuseum Challenge and compare deep learning to \textit{Mensink et al.} \cite{mensink2014rijksmuseum} and other state-of-the-art approaches. We also provide baseline results on four tasks on the new OmniArt data set. After evaluating the models and choosing the one that has the best predictive performance we continue to Stage 2. 

Stage 2 of the experimental setup focuses on evaluating the multi-task model with a different sets of hyper-parameters, data set splits and shared representation sizes against the best performing single task deep learning model. Within this stage we also generate our final results.

\subsection{Datasets}
In our research we rely on multiple data sources like museums, art wiki-sites and pre-compiled datasets. First we crawled a data set from WikiArt containing 126,078 images from more than 3000 individual artists, 150 types and 14 different periods in history. This data set has been used to evaluate various algorithms as it associates artworks with style and genre \cite{bar2014classification, elgammal2015quantifying, karayev2013recognizing, saleh2015large}. However, WikiArt lacks the material information and it is not included in OmniArt at the moment. A relatively new addition to digital artistic data is the on-line collection of the Metropolitan museum with almost half a million artworks combined with extensive meta-data. While assembling this data set, we noticed that almost half of the samples do not belong to the public domain or have no photographic reproduction and cannot be used. A rather smaller data set containing only paintings, is the YourPaintings data set with 5000 painting samples and is publicly available with object level annotations \cite{crowley2014search, kaufmann2004toward}. Featuring 4,000 samples, the French National Library collection is also available \cite{simon2014europeanatech}, but due to lack of material information and its reliance on French annotations, like the YourPaintings data set, it is also excluded at this time.

All results apply to the same datasets and split types. We used 70\% of the data set for training, 20\% for validation and 10\% for testing purposes. The splits were made per class, such that each class had the same distribution during all the experimental phases. 

Our method relies on a single pass through the data set, so it can only be split with respect to a single task 
\begin{math}
(T_{i})
\end{math}
. In our case that task is \textbf{artist attribution} due to the highest number of categories in the data set and the fine-grained nature. Doing a split like this results in an imbalance to the sample distributions per class in the other tasks, but we combat that by assigning task weights
\begin{math}
(w_{i})
\end{math}
using their ratio in the combined loss on the validation set. 

As the artist is the most specific class in the hierarchy, we compute the data distribution with respect to this task. For this reason we can only compare our experimental results to the Rijksmuseum challenge in 2014, on the full data set in period, material and type prediction. 

Using our method, we report baseline performance on the same four tasks on the new OmniArt Challenge as well. We will release the model for feature extraction, data splits and evaluation engine as a museum-centric challenge upon acceptance and continue to gather more data.

\subsubsection{\textbf{The Rijks'14 dataset}}

The Rijks'14 data set \footnote{https://staff.fnwi.uva.nl/t.e.j.mensink/uva12/rijks/} was introduced by \textit{T. Mensink et al.} \cite{mensink2014rijksmuseum} in 2014 as part of the proposed Rijksmuseum challenge. This data set consists of 112,039 photographic reproductions of the artworks exhibited in the Rijksmuseum in Amsterdam. Since the data set for the Rijksmuseum Challenge 2014 is posed from a museum-centric point of view, it offers various object types including paintings, photographs, ceramics, furniture, sculptures, \textit{etc}. Every entry in the data set has at most four labels. If a label is missing or unknown, the entry is assigned an \textit{Unknown} class.

\subsubsection{\textbf{OmniArt Challenge 2017}}

Since 2014, the Rijksmuseum has updated their digitally available content with more than 90,000 photographic reproductions of artworks from their collection. The Metropolitan Museum of Art implemented a new policy known as Open Access, which makes images of artworks it believes to be in the public domain widely and freely available for unrestricted use in accordance with the Creative Commons Zero (CC0) designation. The Met also made meta-data from the entire on-line collection as annotation to the images like artist, title, period, medium and materials and dimensions available on their website. The current estimate of the total number of artworks in their collection is 442,554, but only half of those have photographic reproductions that belong to the public domain. Similar annotations can be found in the Web Gallery of Art data set where 40,000 (c. 28,000 paintings) artworks have been associated with rich meta-data like artists, techniques, period, type, school, geographical origins, etc.

\begin{table}
  \caption{Featured datasets task-wise statistics}
  \label{tab:datasets}
  \begin{tabular}{lccccc}
    \toprule
    Data set  & Entries & Artists  & Types    & Periods    & Materials  \\
    \midrule
    Rijks'14 &  112,039 & 6,626    & 1,054    & 628        & 406        \\
    The Met  &  201,953 & 6,602    & 561      & 2,340      & 5,221		\\
    OmniArt  &  432,217 & 21,364   & 837      & 2,389      & 6,385		\\
    \bottomrule
  \end{tabular}
\end{table}

Using the updated Rijksmuseum collection, the newly available collection from the Met and the Web Gallery of Art collection, we created a new data set containing 432,217 photographic reproductions of artworks combined with rich meta-data. Moreover the quality of annotations is also improved as all the types and materials have been translated to English. Ambiguous labels on the artworks like \textit{Unknown}, \textit{Anonymous} and acronyms have been eliminated in the predefined train, validation and test splits so that the classification problem is well defined. In addition to the previously available data, we offer a meta-data expansion with attributes like \textit{IconClass} \cite{couprie1983iconclass}, \textit{ColorCodes}, \textit{current location}, \textit{real size} and \textit{geographic origins} and \textit{Techniques}.

\subsection{Preprocessing}

Models with deep architectures have the downside of requiring vast amounts of data in order to train properly and learn relevant features \cite{krizhevsky2012imagenet}. Having this in mind, we applied data augmentation techniques to our data to both expand the data set and introduce label-safe variations. 

We experimented with horizontal flips, random rotations, mean subtraction and ZCA whitening. Mean subtraction made performance worse in all cases. While maybe not expected, it is logical for period, type and material prediction, since the integrity of the input sample is important. Subtracting the mean from a metal engraved plate would result in a vague impression of the original image, loosing important texture information. 

We obtained best results when only horizontal flips were applied to random images in the data set which is therefore the only augmentation we used. 

\begin{table*}[]
\centering
\caption{Predictive performance comparison on all four tasks on the Rijks'14 data set}
\label{tab:rijks14_perf}
\begin{tabular}{lcccccccc}
    \toprule
                           & \multicolumn{5}{c}{Artist attribution}                                                    & Type prediction      & Material prediction      & Period estimation                   \\
Metric                     & \multicolumn{5}{c}{\textit{Accuracy (\%)}}                                                & \textit{iMAP (\%)}   & \textit{iMAP (\%)}       & \textit{Mean Abs. Error (years)}    \\\midrule
Targets (\#)               & 374+u          & 374               & 300             & 200             & 100              & 103                  & 81                       & N/A                             \\
Data set usage (\%)         & 100\%          & 59.1\%            & 55\%            & 48\%            & 36.8\%          & 100\%                & 100\%                    & 100\%                               \\\midrule

Mensink et al.              & \textbf{52.8}  & 66.5              & 68.7            & 72.1            & 76.3            & 91.4                 & 94.7                     & 72.4                                \\
Deep CNN (single task)     & 50.3           & 63.9              & 69.5            & 72.5            & 76.7             & 91.7                 & 97.2                     & 71.2                                \\
\textbf{OmniArt  (multi-task)}      & 52.2           & \textbf{67.0}     & \textbf{70.8}   & \textbf{74.0}   & \textbf{78.5}    & \textbf{93.7}        & \textbf{98.0}            & \textbf{70.1}                   \\
                                                                                                              \bottomrule
\end{tabular}
\end{table*}

\subsection{Tasks description}

Since we use the results of The Rijksmuseum 2014 Challenge as our primary baseline, below we describe the different tasks proposed in the challenge on which we evaluate our model.

\subsubsection{\textbf{Artist attribution}}

There are more than 21,000 artists in the OmniArt data set with 23 of them having more than 700 artworks in the collection. In the OmniArt data set there are artworks with an unknown artist. The unknown class is marked with a \textit{+u} in Table \ref{tab:rijks14_perf} for the Rijks'14 dataset. Those artworks are excluded from the experiments too because they might belong to an existing category for the OmniArt challenge. 

Artist attribution is a multi-class classification task. The evaluation block for this task contains a softmax layer and class-wise weight matrix for unbalanced data splits (for Rijks'14).

\subsubsection{\textbf{Creation period estimation}}

The artworks range from ancient pre-historical times to the late 19th century. For many of the artworks that date to early historical periods there is no exact creation date known, therefore estimated creation intervals are provided. In these cases we take the mean  value of the interval as the creation date. We consider the creation period estimation as a regression task with Mean Absolute Error (in years) as the metric. We trained a regressor using the features extracted from our shared representation.

\subsubsection{\textbf{Material prediction}}

There are over 6300 materials attributed to each artwork in the OmniArt data set. This task is a multi-label classification problem as each artwork can have one or more materials. Paper is the most common material appearing in 180,387 artworks in the OmniArt challenge. This contributes to assigning the materials task a lower loss weight because of the unbalanced nature of the task.

As materials can often have noisy labels we applied word stemming to each word for a material. Binary cross-entropy was assigned as the loss function for material prediction with a Mean Average Precision as the metric.  

\subsubsection{\textbf{Type prediction}}

Type prediction is a multi-label classification task with 837 different classes in the OmniArt Challenge. The most common type of artwork is \textit{print(s)} with 108,823 occurrences. Interesting type categories with a significant number of examples are \textit{painting} with 36.785 entries and \textit{photograph} with 31,396 entries. For this task we use the same setup as for Material prediction but with a different task weight.

\begin{figure}
  \includegraphics[width=\linewidth]{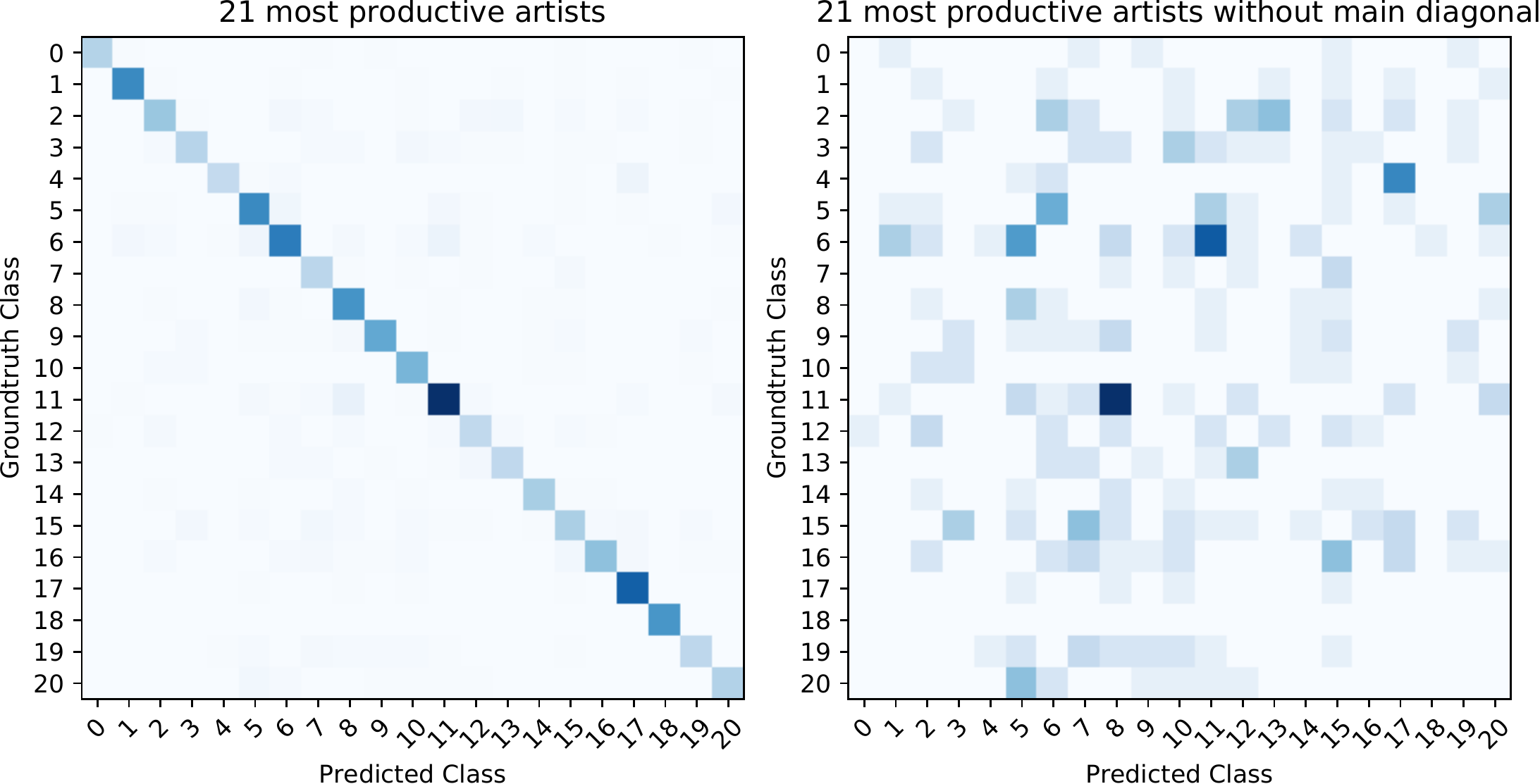}
  \caption{Confusion between the 21 most productive artists (left). The strong main diagonal indicates high classification accuracy. (right) Same confusion matrix with a subtracted main diagonal on the Rijks'14 data set  \label{fig:rijksconf}}
\end{figure}

\subsection{Stage 1: Selecting the Base Layer}

In the first stage we evaluate the performance of popular deep models on each of our four tasks. Evaluation was performed in a fine-tune setting, from train from scratch setting and finally a setting where we use additional external datasets to adjust model weights and filters before fine-tuning for the task at hand. Fine-tuning is performed by training only the last layer of the models and not modifying the filters learned beforehand. This stage allows to see how different model architectures react to each of our tasks and gives insight into which model would perform best for all tasks in a multi-task scenario.

\begin{figure}
  \includegraphics[width=\linewidth]{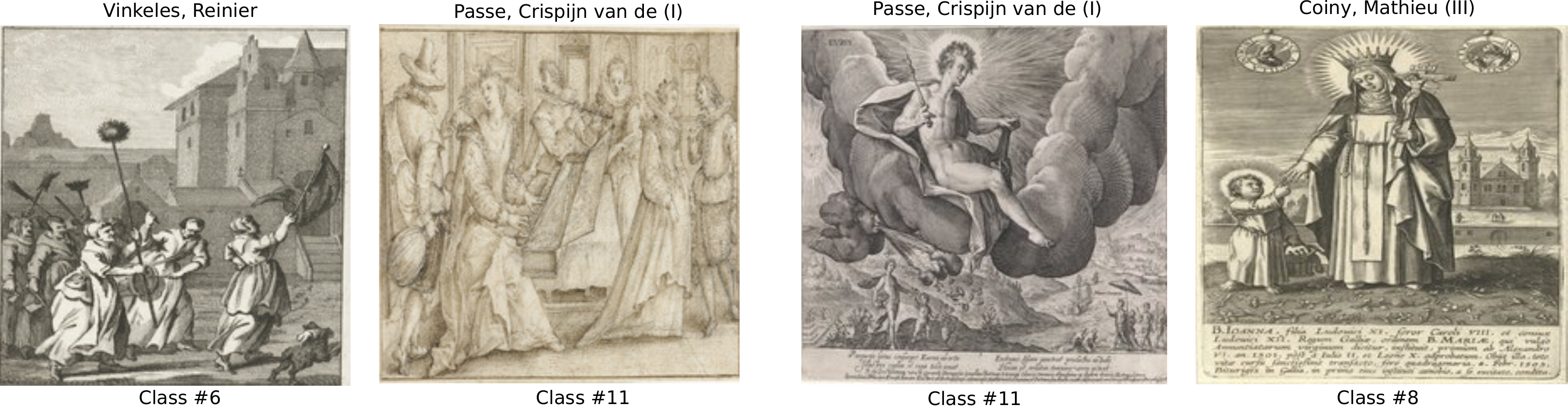}
  \caption{OmniArt's confusion in artist attribution on Rijks'14 \label{fig:rijks14_omniart_conf}}
\end{figure}

Stage 1 of the experimental design is about testing model performance on individual tasks for assessing the best architecture to use for their combination. We experimented with several of the best performing deep architectures on ImageNet like Resnet-50 \cite{he2016deep}, VGG-16, VGG-19 \cite{DBLP:journals/corr/SimonyanZ14a} and Inception v2 \cite{szegedy2015going}. We obtained best results with the features from the ResNet-50 model (without the top block) and continued to use it as the main feature extraction unit in the other experiments.

This stage of our experimental design is especially important because it can be directly compared with the state-of-the-art approaches in all four tasks since we can use the same data splits. Table \ref{tab:rijks14_perf} shows a direct comparison between the handcrafted feature approach from \textit{Mensink et al.}, a CNN and our method OmniArt.

\begin{table}
  \caption{OmniArt artist attribution performance on Rijks'14 \label{tab:rijks14_artist_perf}}
  
  \begin{tabular}{ccccc}
    \toprule
    \# Artists  & Min. Samples/Class & Top-1 Acc (\%) & Top-3 Acc (\%) \\
    \midrule
    6           & 1,100              & 91.9           & 99.6           \\
    22          &   500              & 84.8           & 95.1           \\
    52          &   300              & 81.9           & 92.0           \\
  	186         &   100              & 74.3           & 84.4           \\
    \bottomrule
  \end{tabular}
\end{table}

\begin{table*}[]
\centering
\caption{Predictive performance comparison on all four tasks for OmniArt on 4 splits with respect to the Artist task: a) >100 samples, b) >300 samples, b) >700 samples and  d) >1100 samples per class on the OmniArt data set. }
\label{tab:omniart_perf}
\begin{tabular}{lccccccccccccccccccccc}                                                                                                                                                                                                                     \toprule
             & \multicolumn{4}{c}{Artist attribution}     &&\multicolumn{4}{c} {Type prediction}    && \multicolumn{4}{c} {Material prediction}  && \multicolumn{4}{c} {Period estimation}    \\
Metric       & \multicolumn{4}{c}{\textit{Accuracy (\%)}} &&\multicolumn{4}{c} {\textit{MAP (\%)}}  && \multicolumn{4}{c} {\textit{MAP (\%)}}    && \multicolumn{4}{c} {\textit{Mean Abs. Error (years)}} \\\midrule
Targets (\#) & 390  & 87    & 23   & 8                    && 112  & 75   & 39    & 21               && 1424   & 803    & 94      & 63            && 544   & 510    & 358    & 237                \\\midrule

Deep CNN     & 60.7  & 76.0  & 82.8  & 91.0               && 99.0 & 98.9 & 95.9  & 96.0             && 97.1   & 97.0   & 84.2    & 74.0          && 79.3  & 69.9   & 54.2   & 31.1    \\
\textbf{OmniArt}      & \textbf{64.5 } & \textbf{80.8}  &\textbf{ 87.5}  & \textbf{94.1}               && \textbf{99.4} & \textbf{99.7} & \textbf{98.8 } & \textbf{97.9}             && \textbf{99.0}   & \textbf{98.8}   & \textbf{85.5}    & \textbf{76.8}          && \textbf{77.9}  & \textbf{67.8}   & \textbf{52.2}   & \textbf{28.5}    \\\bottomrule
\end{tabular}
\end{table*}

\subsection{Stage 2: Evaluating our multi-task method}

The second stage consists of constructing the optimal architecture that complies with the multi-task nature of the problem. In the method section we explained that a shared representation can be beneficial to tasks whose targets are correlated. Configuring the optimal shared representation is the goal we tried to accomplish in this stage along with finding the best shared representation size that fits every split of the data set.

In this stage we tested various hyper-parameters and selected the overall best performing setup, as all the tasks have a different nature and are prone to react differently to changes in the architecture. We evaluate our multi-task model by comparing our scores on all 4 tasks proposed in The Rijksmuseum Challenge. For the Rijks'14 data set we compare with \textit{Mensink et al.} \cite{mensink2014rijksmuseum} as they are the original creators of the Rijksmuseum Challenge and have scores for all proposed tasks. For the OmniArt data set, we compare the performance of our method versus a single task deep convolutional neural network. In Table \ref{tab:rijks14_perf} we observe that best performance is obtained on all 4 tasks with the multi-task method. However in the \textit{374+u} case, handcrafted features outperform the deep nets in artist attribution. 

\begin{figure}

  \includegraphics[width=\linewidth]{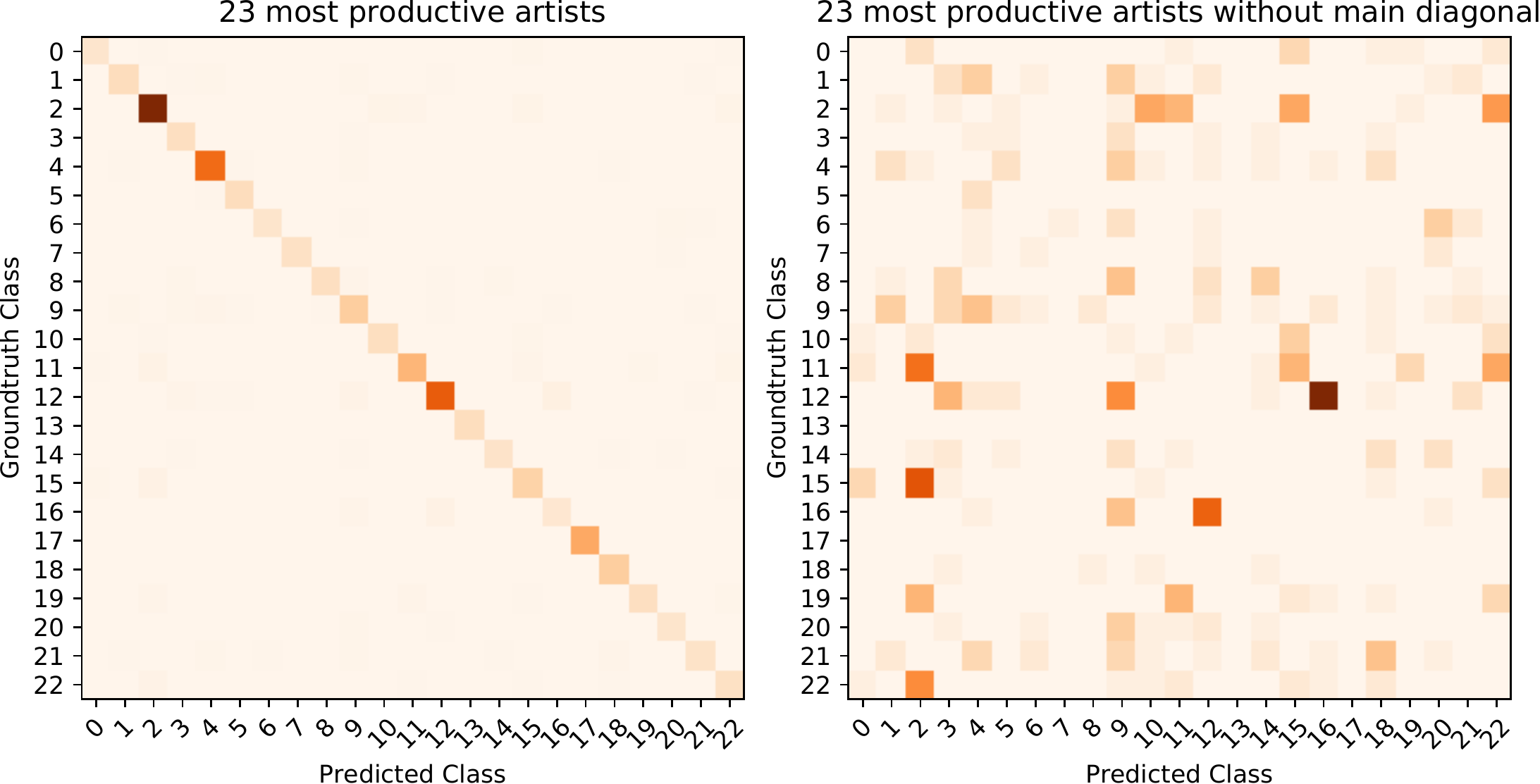}
  \caption{Confusion between the 23 most productive artists (left). (right) Same confusion matrix with a subtracted main diagonal on the OmniArt data set \label{fig:omniconf}}
\end{figure}

This is possibly due to the very limited number of examples per class which does not allow for a good representation to be learned, while the handcrafted features maintain their quality even for such low amounts of data. In Table \ref{tab:omniart_perf} we see the performance of the OmniArt method versus a single task deep CNN. An interesting find is that as samples per artist threshold lowers, artist attribution performance drops but it is the other way around for the multi-label tasks. We believe that as we use a higher percentage of the data set, we get more samples per class in the multi-label setting while the number of output targets remains the same which is important for the representation learning. 

Shared representation size is also important for performance. We experimented with sizes from 1024 units up to 8192 units and achieved best overall balance for all tasks with 6144 units. Through the experiments in this stage we observed that having a smaller shared representation layer is beneficial for period estimation where a mean drop of 8.3 years in absolute error can be achieved with smaller shared representation sizes.

An interesting fact is that while maintaining good discriminative performance, the multi-task method shortens training and testing times significantly making it more efficient than the model-per-task methods. We calculate that it takes OmniArt 6.22 s to go through 200 batches of 32 images for all 4 tasks in the testing phase. A single-task CNN takes 2.13 s per task for the same setup. In this case this is a speed up of 25\%. A rather larger improvement in execution times can be observed in the training phase. OmniArt takes circa 73 minutes to train with the > 1100 samples per class setting on a single Nvidia Titan X, while the combined training times of the four single task models is 198 minutes, which is \textbf{2.6 times} slower than our multi-task method.

\begin{figure}

  \includegraphics[width=\linewidth]{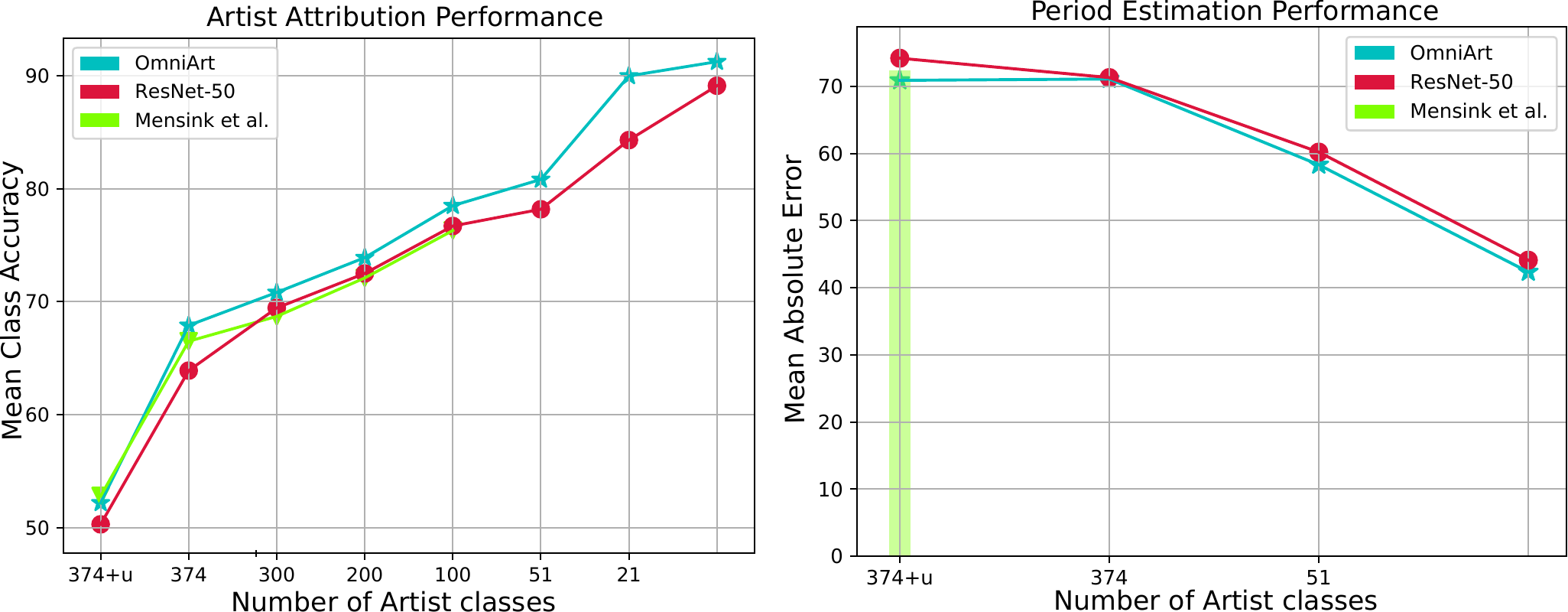}
  \caption{Artist attribution and period estimation performance comparison with a CNN on the Rijks'14 data set  \label{fig:rijks14_artist}}
\end{figure}

\subsection{Qualitative analysis}

 Quantitative performance measures show good artist attribution performance, however there are misclassifications in artist attribution. Further exploration of the inner confusion between classes clearly visible in Figure \ref{fig:omniconf} and Figure \ref{fig:rijksconf} after we remove the main diagonal, revealed an interesting find we call \textbf{The Luyken case}. 

 \textbf{The Luyken case} originates from the confusion between late 17th century Dutch book illustrators Caspar Luyken (Class 16) and Jan Luyken (Class 12) clearly visible in Figure \ref{fig:omniconf}. Caspar Luyken was the son of Jan Luyken (a Dutch book illustrator) \cite{franits2004dutch}. His father was also his teacher and they often worked together using the same techniques and materials. Being a father and son, they also share the same period of existence, which with a 52.2 years of mean error margin can easily be missed. However, almost all of the confused samples are from their mutual active period, while the correct ones are either from the later stage of the son's productive live or the early stage of the father's. Additional dataset exploration shows that the correctly specified artworks have a difference in material and period, which furhter demonstrates the benefits of viewing this as a multitask problem. This find would suggest an evolution in technique of the son after separating from the father and leaving to work for an art dealer in Germany \cite{doi:10.7227/ALX.21.2.2}. Such results speak in favor to our hypothesis that semantic entanglement between different data attributes in separate tasks can be modeled in a shared representation with our method. 



\section{Conclusions}

The OmniArt method outperforms the current state-of-the-art approaches on the Rijks'14 datasets and speeds up training and testing times. In terms of best quantitative results, the performance can benefit from conventional boosting techniques like training ensembles of models, applying different voting techniques and tweaking hyper-parameters. While absolute numbers are an important quality and performance measure, the base improvement is sufficient to prove our method as a valid one. The OmniArt challenge which we continue to expand and improve, is presented in the form of a challenge to stimulate further research and development in the artistic data domain. While this work focuses solely on artistic data, the same idea can be easily adapted and transfered to different domains. 

\bibliographystyle{ACM-Reference-Format}
\bibliography{sigproc} 

\end{document}